\documentclass{fmj2010}
\begin{document}
   \title{Rapid Variability of Gamma-ray Emission from Sites near the
43 GHz Cores of Blazar Jets}
\titlerunning{Rapid Gamma-ray Variability near the 43 GHz Core}

   \author{A.\ P.\ Marscher and S.\ G.\ Jorstad}

   \institute{Institute for Astrophysical Research, Boston University, USA
             }

\abstract{ Comprehensive VLBI and multi-waveband monitoring indicate that a single
superluminal knot can cause a number of $\gamma$-ray flares at different
locations. However, the often very rapid variability timescale is a challenge to
theoretical models when a given flare (perhaps the majority of those observed) is inferred from
observations to lie near the 43 GHz core, parsecs from the central engine. We present
some relevant observational results, using the BL Lac object AO~0235+164 as an example. We
propose a turbulent cell model leading to a frequency-dependent filling factor of the
emission region. This feature of the model can provide a solution
to the timescale dilemma and other characteristics of blazar emission.   }

   \maketitle
%

\section{Introduction}

There exists a crisis in the interpretation the results of
multi-waveband monitoring of blazars. Comparisons of dates of peak $\gamma$-ray flux
measured by EGRET with epochs of ejections of superluminal radio knots (Jorstad et al.\ 
\cite{jor01a}), onsets of millimeter-wave flares (L\"ahteenm\"aki \& Valtaoja
\cite{laht03}), and changes in centimeter-wave polarization (Jorstad et al.\ 
\cite{jor01b}) lead to the conclusion that most $\gamma$-rays outbursts
are coincident with radio events. Yet we know that the jets of blazars are opaque to
radio emission within $\sim 1$ pc or more from the central engine. This is difficult
to reconcile with the shortest time scales of GeV variability observed --- a few hours
in PKS~1622$-$297 (Mattox et al.\ \cite{mat97}) and 3C~454.3
(Foschini et al.\ \cite{fos10}).

However time scales of variability reflect
size scales, not distance from the central engine. Furthermore, jets are very narrow,
with opening half-angle $\sim 10^\circ/\Gamma$, where $\Gamma$ is the bulk Lorentz
factor of the flow in the jet (Jorstad et al. \cite{jor05}).
Because of this, a jet can have a
cross-sectional radius much less than its distance from the central engine. Nevertheless,
this does not solve the entire problem. The core on 43 GHz Very Long Baseline Array
(VLBA) images of PKS~1510$-$089 has been estimated to lie $\sim 20$ pc from the central
engine (Marscher et al.\ \cite{mar10a}). Offsets of parsecs are implied for other luminous
blazars as well (e.g., Marscher et al.\ \cite{mar08}; Chatterjee et al.\ \cite{chat08}).
The cross-sectional radius of a luminous
blazar is $\sim 0.1$ pc near the 43 GHz core. With a Doppler factor $\delta \sim 20$ and
redshift $\sim 0.5$, the shortest time scale of variability should be $\sim 1$ week, not
a fraction of a day.

Because of this discrepancy, a number of authors have insisted that the $\gamma$-ray
emitting region lies within $\sim 10^{16}$ cm of the central engine (e.g.,
Tavecchio et al.\ \cite{tav10}).
This has the advantage for high-energy emission models that optical-uv photons from
the broad emission-line region are available for scattering to $\gamma$-ray energies by
highly relativistic electrons in the jet. Poutanen \& Stern
(\cite{ps10}) find that pair production off the
uv photon field can then cause the observed sharp break in the spectrum of some blazars
at GeV energies. However, such an interpretation requires that
the timing of the radio and $\gamma$-ray events is a chance coincidence in every case.
This seems unlikely, especially with singular events such as the ultra-high-amplitude
optical flare, very sharp $\gamma$-ray flare, and passage of a superluminal knot through
the 43 GHz core in PKS~1510$-$089 at essentially the same time (Marscher et al.\
\cite{mar10a,mar10b}). Furthermore, the number of $\gamma$-ray outbursts observed
with {\it Fermi} that are either coincident with or follow the passage
of a new superluminal knot through the centroid of the core is becoming high
enough to conclude that a large fraction of the $\gamma$-ray emission originates parsecs
away from the central engine (Jorstad et al. \cite{jor10} and these proceedings).

We propose to solve this dilemma by developing a model in which much of the optical and
high-energy radiation in a blazar is emitted near the 43 GHz core in VLBA
images, parsecs from the central engine. The model allows for short time scales of
optical and $\gamma$-ray variability by restricting the highest-energy electrons radiating
at these frequencies to small sub-regions of the jet. That is, in our model the filling
factor at high frequencies is relatively low, while that of the electrons radiating at
$\sim 10^{10-13}$ Hz is near unity. Such a model is consistent with other prominent
features of optical vs.\ lower frequency emission.

\section{Properties of blazar emission to incorporate in the development of a model}
   \begin{figure}
   \centering
   \vspace{8cm}
   \includegraphics{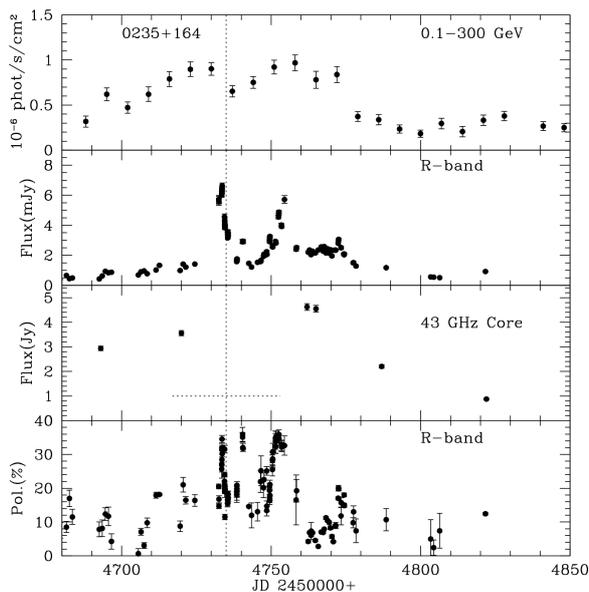}
      \caption{Multi-waveband light curves of the BL Lac object AO~0235+164 in late 2008.
           The time when the superluminal knot (see Fig. 2) passed through the core is marked
           by a vertical line, with the horizontal line in panel 3 indicating the uncertainty
           in the date of this event.}
         \label{fig1}
   \end{figure}

\begin{figure}
   \centering
   \vspace{4.2cm}
   \includegraphics{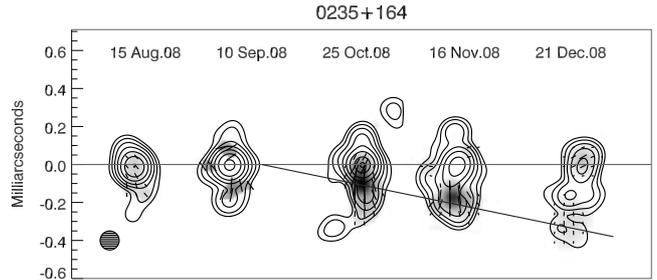}
      \caption{VLBA images at 43 GHz of 0235+164 in late 2008. Contours (factors of 2 starting
         at 0.023 Jy/beam) correspond to
         total intensity and gray scale to polarized intensity. Line segments indicate
         direction of polarization electric vectors. A circular Gaussian
         restoring beam is used with FWHM similar to the resolution of the longest VLBA baselines.
         Motion of the superluminal knot is indicated by the diagonal line.}
         \label{fig2}
   \end{figure}

\begin{figure}
   \centering
   \vspace{11cm}
   \includegraphics{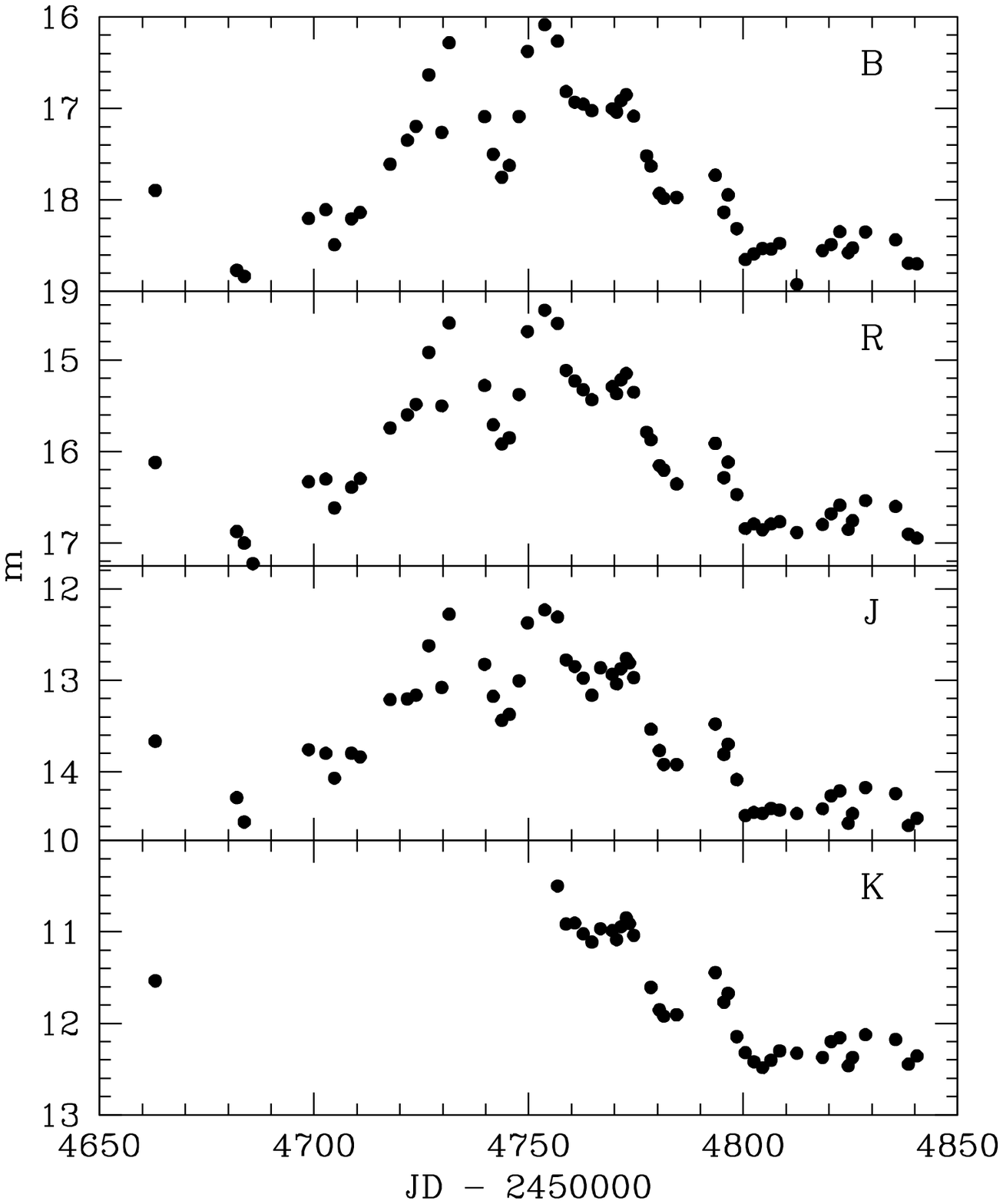}
   \includegraphics{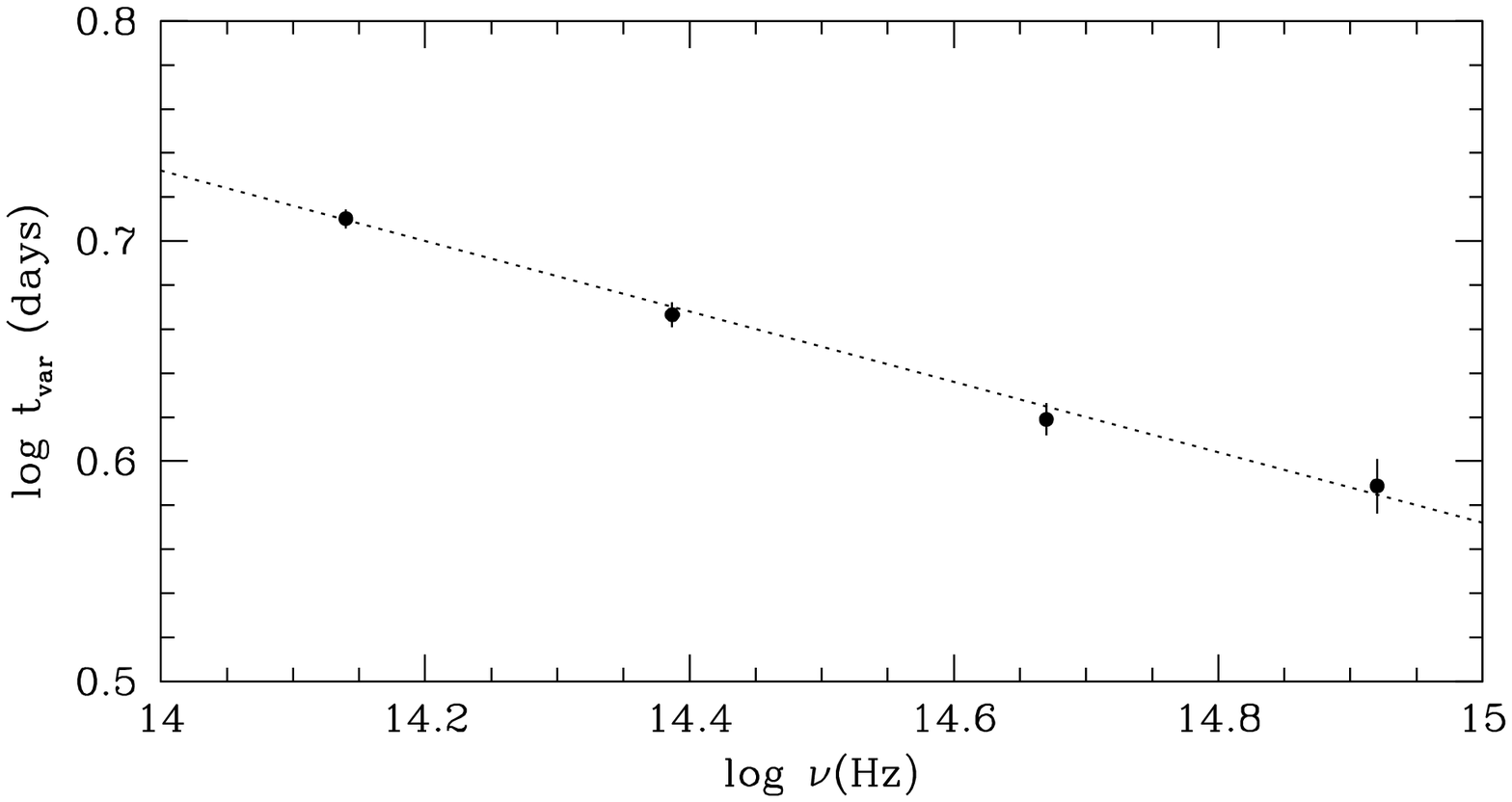}
      \caption{{\it Top:} Multi-band optical/near-IR light curves of 0235+164 during the interval
         displayed in Fig.\ \ref{fig1}. Magnitudes here are not corrected for extinction.
         {\it Bottom:} Time scale of flux variability (as defined in the text) as a function of
         optical/near-IR frequency of 0235+164 between Julian dates 2454756.7 and 2454758.7.
         Diagonal line represents a power law with slope $-0.16$. Data are from the
         Yale Fermi/SMARTS project, website http://www.astro.yale.edu/smarts/glast/.}
               
         \label{fig3}
   \end{figure}

We now have at our avail light curves at many frequencies across the electromagnetic
spectrum, polarization vs.\ time data at radio, mm-wave, and optical/near-IR bands, and
well-sampled sequences of VLBI images---in both total and polarized intensity---at high
radio frequencies for many blazars. This is providing a wealth of information that must
be considered when developing models for the nonthermal emission from blazar jets.
This information includes the following:\\
  \\
(1) In most blazars, the detailed properties of the synchrotron emission change gradually as
the wavelength of observation decreases from the millimeter to optical range. (a) The spectral
index ($\alpha \equiv - d\;log F_\nu/d\;log\,\nu$) steepens from a value usually in the range of 
0.5-1.0 to $> 1$; (b) the linear polarization increases on average and becomes more highly
variable (D'Arcangelo et al.\ \cite{darc09});
and (c) the time scale of flux variability, [$t_{\rm var} \equiv (t_2-t_1)/\vert ln(F_2/F_1) \vert$,
where $t_1$ and $t_2$ are times when flux densities $F_1$ and $F_2$ are measured, respectively]
decreases. An example of the latter is the BL Lac object AO~0235+164 ($z = 0.94$) during
a multi-waveband outburst in 2008 (Fig.\ \ref{fig1}). As seen in
Figure \ref{fig2}, a bright, extremely superluminal ($\sim 70c$!)
knot passed through the core at 
essentially the same time as the peak of a sharp optical flare during a $\sim 80$-day
$\gamma$-ray outburst. During the declining phase of another major flare observed at
two optical and two near-IR bands, the shortest measured time scale of flux decrease depended on
frequency $\nu$ as $t_{\rm var} \propto \nu^{-0.16}$ (see Fig.\ \ref{fig3}).
  \\ 
  \\
(2) The period of elevated optical flux in 0235+164 (see Fig.\ \ref{fig1})
matches that of the high $\gamma$-ray
state, but there were no high-amplitude $\gamma$-ray flares during the two main
optical events. Yet electrons with sufficient energies to emit optical synchrotron radiation
for typical magnetic field strengths of $\sim 0.5$ G (Hagen-Thorn et al.\ \cite{ht08}),
$E \sim 10^4 mc^2$,
should also be able to scatter those photons (or other optical photons from the surroundings)
to $\gamma$-ray energies.\
  \\
(3) The power spectral density (PSD) of flux variations of the $\gamma$-ray bright blazar
3C~279 follows a power law (Chatterjee et al.\ \cite{chat08})
without any indication of a break on time scales from $\sim 1$ day to 12 years.\\
  \\
(4) The 43 GHz core on VLBI images lies at a significant distance from the central engine.
This must be the case theoretically based on the self-absorption turnover in the synchrotron $F_\nu$
vs.\ $\nu$ spectrum at millimeter wavelengths (Impey \& Neugebauer \cite{in88}). If emission
in the jet continued
in a self-similar way down to a size less than about $10^{15}$ cm, the turnover would occur at
much shorter wavelengths. In the radio galaxy 3C~120, time delays between a drop in X-ray flux
from the central engine and the passage of a new superluminal knot through the core at 43 GHz
led Marscher et al.\ (\cite{mar02}) and Chatterjee et al.\ (\cite{chat09}) to conclude that the
core lies 0.5 pc from the
black hole. In a more powerful source, such as the quasars 3C~279 and PKS~1510$-$089, timing
studies indicate that the core lies tens of parsecs (after de-projection) from the central engine
(Marscher et al. \cite{mar10a,mar10b}; Chatterjee et al.\cite{chat08}).

\section{A turbulent cell model for variability in blazars}

The red-noise behavior of the PSD of 3C~279 (Chatterjee et al.\ \cite{chat08})
implies that the variations
are stochastic in nature. If main events in the light curves associated with bright superluminal
knots are caused by shocks, these shocks must be produced randomly with a range of compression
ratios. The short-term fluctuations can be understood as the consequence of
turbulent ambient jet plasma that passes through the shocks (Marscher, Gear, \& Travis
\cite{mgt92}). If this is the case, then the multi-wavelength behavior of the emission depends
on the range of physical parameters of the turbulence.

We propose a model in which the emission arises from a collection of turbulent cells of similar
size, each with strength of the magnetic field $B$ (with random orientation) and density of
electrons $n_{\rm e}$ that follow a probability distribution, with higher values less likely. We 
imagine that the turbulent cells pass into and out of a disk-shaped primary emission region
representing a shock, bounded by the shock front and, probably, a rarefaction.
Electrons are injected with a power-law energy distribution at the shock front, after which
the cell advects away from the shock. The cooling time of the electrons depends on $B$.
If inverse Compton losses are important, then it will also depend on the energy density of
photons $n_\gamma$; we defer consideration of this case to a subsequent study.

The key addition to this model over that proposed in Marscher et al.\ (\cite{mgt92}) is that
the maximum electron energy $E_{\rm max}$ varies from cell to cell.
According to models of energization of particles at 
relativistic shock fronts (e.g., Baring \cite{bar10}),
the efficiency of particle acceleration depends inversely on the ratio of the gyro-radius to
mean free path for pitch-angle scattering. This ratio should be larger when $B$
is stronger, so that the amplitude of the magnetic turbulence rises. If the
value of $B$ in a cell follows a power-law probability distribution above some
minimum value, then we might expect $E_{\rm max}$ to do the same. (Or the slope of
the resultant electron energy distribution might be flatter for stronger fields, thereby
increasing the relative number of electrons with the highest energies. This would produce a
similar effect on the synchrotron spectrum as the case considered here.)

\begin{figure}
   \centering
   \vspace{6.5cm}
   \includegraphics{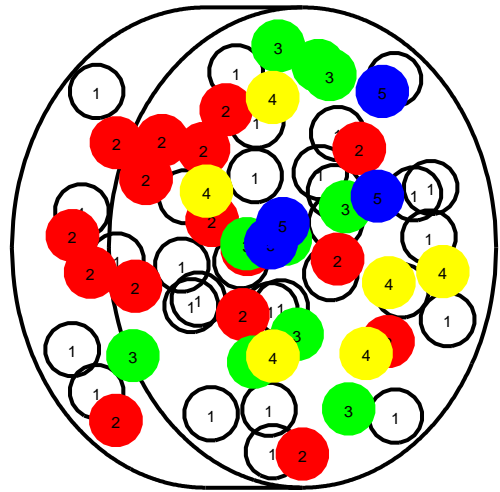}
      \caption{Sketch of our proposed model consisting of a number of emission cells
               (circles) inside a disk (representing, for example, a shock wave),
               each with a different maximum energy of relativistic
               electrons. All numbered cells have electrons with sufficiently high energies to
               emit synchrotron radiation at a relatively low frequency, e.g., $< 10^{13}$
               Hz. Only the cells marked 2 and higher have electron energies high enough to
               emit at $> 10^{13.5}$ Hz. Only those marked 3, 4, and 5
               can emit at $> 10^{14}$ Hz, only those numbered 4 and 5
               can emit at $> 10^{14.5}$ Hz, and only those numbered 5
               can emit at $> 10^{15}$ Hz.}
         \label{fig4}
   \end{figure}

The different values of $B$, $E_{\rm max}$, and $n_{\rm e}$ in different cells, with higher
values occurring more rarely, leads to a frequency dependence of the synchrotron emission.
Recall that, in
the scenario envisioned by Marscher \& Gear (\cite{mg85}), the synchrotron spectrum steepens
by 1/2 because the dimension transverse to the shock front, and therefore the volume
$V$, of the emitting region follows $V(\nu) \propto \nu^{-1/2}$. This will still apply at
sufficiently high frequencies if electrons are injected only
as a cell passes across the shock front. However, we introduce an additional frequency
dependence on the volume because only some fraction of the cells contain electrons
that can radiate at frequency $\nu$ after they cross the shock front. We illustrate the
geometry of our model in Figure \ref{fig4}.

We have not yet developed a definite relationship between the electron energy distribution
and the power spectrum of the turbulence. Our model therefore does not at this stage predict
the slope of the spectrum at frequencies where $V(\nu) < 1$. We can, however, turn to
observations to infer the required functional form of $V(\nu)$. The relationship
$t_{\rm var} \propto \nu^{-0.16}$ found above for 0235+164 could correspond to stochastic
variations in $N$ cells if $N(\nu) \propto V(\nu) \propto \nu^{-0.32}$, since we expect that the level
of fluctuations in flux $\langle \Delta F/F \rangle \propto N^{-1/2}$. This is in agreement
with the steepening of the continuum spectrum from far-IR ($\alpha \approx 1.0$) to
optical (($\alpha \approx 1.3$) at times when there is no flare. Since 0.32 is less than
0.5, we infer that cells can cross the entire emission region before electrons that radiate
at B band suffer significant synchrotron losses in 0235+164.

   \begin{figure}
   \centering
   \vspace{4.4cm}
   \includegraphics{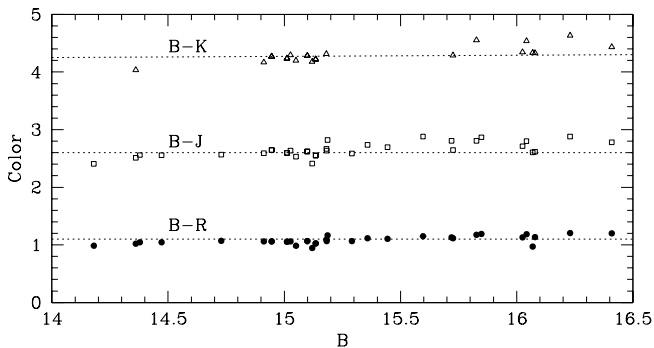}
      \caption{Optical/near-IR colors vs.\ B magnitude of 0235+164 during the interval
         covered in Fig.\ \ref{fig1}. All magnitudes and colors have been corrected for
         foreground reddening in both our Galaxy and an intervening galaxy at redshift
         0.524 according to
         Table 5 of \cite{rait05}. Comparison with the horizontal dotted lines shows
         the trend of flatter spectra as the flux increases. For example, the $B-J$ color
         difference between 2.4 at $B=14.17$ and 2.85 at $B=2.85$ corresponds to a change
         in the spectral index from $\alpha = 1.01$ to 1.34. Data source is the same as
         for Fig.\ \ref{fig3}.}
               
         \label{fig5}
   \end{figure}

Since a rapid flare corresponds to one or more new cells with higher than average $B$ or
$n_{\rm e}$, the spectrum should flatten near the peak of the flare, since
the volume of the enhanced emission is roughly the same at all wavelengths within a 
relatively narrow range (e.g., optical/near-IR). As Figure
\ref{fig5} demonstrates, this indeed occurred during the 80-day $\gamma$-ray outburst of
0235+164. The change in spectral index closely matches that required by the model
($\Delta \alpha = 0.32$).

The higher mean level and amplitude of fluctuations of linear polarization at higher
frequencies is a natural consequence of the turbulent cell model. In the extreme case
of either very little compression at the electron injection front (weak shock) or a shock
viewed almost exactly face-on (so that the component of $\bf B$ in the plane of the sky
remains chaotic),
the mean degree of polarization at frequency $\nu$ would be
$\sim 0.75 N(\nu)^{-1/2}$ (Burn \cite{burn66}) and the standard deviation $\sim 0.75[2N(\nu)]^{-1/2}$
(Jones \cite{jones88}). This becomes more complicated when the compression is substantial and
the front lies at an angle to the sky plane. There will still be frequency-dependent
polarization, but simulations are needed for a more quantitative analysis.

The time scale of variability $t_{\rm var} \propto \nu^{-0.16}$ in 0235+164 at optical/near-IR
frequencies, probably extending down to $\sim 10^{13}$ Hz.
If we assume that the volume of emission at this
frequency includes the entire superluminal knot, whose angular diameter was 0.05
mas on JD 2454762 as measured by modeling the feature as a face-on disk and fitting
the VLBA data. This corresponds to a radius of 0.2 pc. If we adopt $\delta \approx 70$, the typical
time scale of variability at $\nu \sim 10^{13}$ Hz should be $\sim 7$ days. This becomes
$\sim 3$ days at B band, close to the observed value of 4 days (see Fig.\ \ref{fig3}).

Inverse Compton emission depends on $n_{\rm e}$ and $n_\gamma$,
while the synchrotron flux depends on the former and on the strength and direction of the
local magnetic field. We therefore expect correlations between the optical and $\gamma$-ray
emission, but not exact correspondence of light curves.
The geometrical break in the synchrotron spectrum should appear also in the $\gamma$-ray
spectrum from inverse Compton scattering. We plan to investigate this possibility alongside
the detailed time behavior of the multi-waveband emission through numerical simulations of
the model in the near future.

\begin{acknowledgements}
This research is supported in part by NNX08AV65G, NNX08AV61G, and NNX09AT99G,
and National Science Foundation grant AST-0907893.
The VLBA is an instrument of the National Radio Astronomy Observatory, a facility of the NSF,
operated under cooperative agreement by Associated Universities, Inc.

\end{acknowledgements}

\end{document}